# A Unifying Framework for Local Throughput in Wireless Networks

Pedro C. Pinto, *Student Member, IEEE*, and Moe Z. Win, *Fellow, IEEE*

*Abstract*—With the increased competition for the electromagnetic spectrum, it is important to characterize the impact of interference in the performance of a wireless network, which is traditionally measured by its *throughput*. This paper presents a unifying framework for characterizing the local throughput in wireless networks. We first analyze the throughput of a probe link from a *connectivity perspective*, in which a packet is successfully received if it does not collide with other packets from nodes within its reach (called the audible interferers). We then characterize the throughput from a *signal-to-interference-plus-noise ratio (SINR) perspective*, in which a packet is successfully received if the SINR exceeds some threshold, considering the interference from all emitting nodes in the network. Our main contribution is to generalize and unify various results scattered throughout the literature. In particular, the proposed framework encompasses arbitrary wireless propagation effects (e.g, Nakagami-$m$ fading, Rician fading, or log-normal shadowing), as well as arbitrary traffic patterns (e.g., slotted-synchronous, slotted-asynchronous, or exponential-interarrivals traffic), allowing us to draw more general conclusions about network performance than previously available in the literature.

## I. INTRODUCTION

A wireless network is typically composed of many spatially scattered nodes, which compete for shared network resources, such as the electromagnetic spectrum. A traditional measure of how much traffic can be delivered by such a network is the packet throughput.[1] In a wireless environment, the throughput is constrained by various impairments that affect communication between nodes, namely: the *wireless propagation effects*, such as path loss, multipath fading, and shadowing; the *network interference*, due to signals radiated by other transmitters; and the *thermal noise*, introduced by the receiver electronics. It is therefore of interest to develop a framework that quantifies the impact of all these impairments on the throughput of the network. Such framework should also incorporate other important network parameters, such as the spatial distribution of nodes and their transmission characteristics. We accomplish such goal by using fundamental tools from stochastic geometry.[2]

The performance analysis of MAC protocols has received tremendous attention in the last four decades, from the throughput of legacy ALOHA protocols [3], to the capture effect in ALOHA systems [4]–[7] and in IEEE 802.11 (CSMA/CA) systems [8], [9]. In this context, the *spatial distribution of nodes* in the network plays a key role in determining the signal and interference levels at each receiver, thus affecting the connectivity and throughput of the network.

In the topic of connectivity of spatially distributed networks, the probability of node isolation in a spatial Poisson process is derived in [10] for the case of log-normal shadowing, and in [11] for combined Rayleigh fading and log-normal shadowing. The distribution of the number of nodes which can communicate with one another in a spatial Poisson process is analyzed in [12]–[16], assuming log-normal shadowing. In the topic of throughput of spatially distributed networks, the throughput of ALOHA channels for networks distributed in space is considered in [3], [4], [17], [18], but ignoring wireless propagation effects such as fading or shadowing. The local throughput in a cellular packet network is analyzed in [19]–[21], for some particular combination of the location of nodes, channel access, and wireless propagation characteristics. Other issues studied in the literature include the scaling behaviour of the network capacity [22]–[24], the distribution of the aggregate interference power [25], the characteristic function of the aggregate interference [26], and the moments of the aggregate interference [27], [28]. However, such literature is largely constrained to some restrictive combination of path loss exponent, propagation model, spatial configuration of nodes, and packet traffic. Furthermore, the existing results are not easily generalizable if some of these system parameters are changed.

In this paper, we develop a unifying framework for the characterization of connectivity and throughput in wireless networks, where the nodes are scattered according to a spatial Poisson process. The main contributions of the paper are as follows:

- *Unifying framework for connectivity and throughput:* The proposed framework generalizes and unifies various results in the literature, by accommodating arbitrary

P. C. Pinto and M. Z. Win are with the Laboratory for Information and Decision Systems (LIDS), Massachusetts Institute of Technology, Room 32-D674, 77 Massachusetts Avenue, Cambridge, MA 02139, USA (e-mail: ppinto@mit.edu, moewin@mit.edu).

This research was supported, in part, by the Portuguese Science and Technology Foundation under grant SFRH-BD-17388-2004; the MIT Institute for Soldier Nanotechnologies; the Office of Naval Research under Presidential Early Career Award for Scientists and Engineers (PECASE) N00014-09-1-0435; and the National Science Foundation under grant ECS-0636519. This work was presented, in part, at the IEEE Military Communications Conference, San Diego, CA, Nov. 2008, and at the IEEE Global Telecommunications Conference, Honolulu, HI, Nov. 2009.

[1]In this paper, the term *throughput* refers to *local throughput* or *one-hop throughput*. As defined later, it corresponds to the probability of a successful one-hop transmission, and is a useful metric in the analysis and design of medium access control (MAC) protocols. The throughput can also be considered from an information theoretic perspective, where it is defined as the rate supported with reliability from source to destination over multiple hops.

[2]An overview on the application of stochastic geometry to various problems in wireless communications can be found in the tutorial papers [1], [2].

wireless propagation effects as well as arbitrary traffic patterns.

- *Connectivity in arbitrary wireless propagation environments:* We provide a probabilistic characterization of the number of audible nodes in a spatial Poisson process, using fundamental tools of stochastic geometry. Such characterization is more general than in the previous literature, since it is valid regardless of the considered propagation scenario.
- *SINR in arbitrary propagation characteristics and packet traffics:* We provide a probabilistic characterization of the SINR of a link subject to the aggregate interference and noise. Such general characterization is valid regardless of the considered type of propagation scenario or packet traffic.
- *Throughput in connectivity-based and SINR-based scenarios:* We obtain expressions for the throughput of a link, both from both a connectivity perspective and an SINR perspective. Due to the generality of the propagation and packet traffic models, we are able to draw more general conclusions than previously available in the literature.

This paper is organized as follows. Section II describes the system model. Section III characterizes the throughput of the probe link in a connectivity-based approach. Section IV characterizes the throughput in a SINR-based approach. Section V provides numerical results to illustrate the dependence of the throughput on important network parameters. Section VI concludes the paper and summarizes important findings.

## II. SYSTEM MODEL

### A. Spatial Distribution of Nodes

Following [1], we model the spatial distribution of the nodes according to a homogeneous Poisson point process in the two-dimensional infinite plane, with a spatial density $\lambda$ (in nodes per unit area). This spatial model is depicted in Fig. 1. For analytical purposes, we assume there is a *probe link* composed of two nodes: one located at the origin of the two-dimensional plane (without loss of generality), and another one (node $i = 0$) deterministically located at a distance $r_0$ from the origin.[3] All the other nodes ($i = 1 \ldots \infty$) are interfering nodes, whose random distances to the origin are denoted by $\{R_i\}$, where $R_1 \leq R_2 \leq \ldots$. Our goal is then to determine the throughput of the probe link subject to the effect of all the interfering nodes in the network.

### B. Transmission Characteristics of Nodes

We consider interfering nodes with the same transmit power $P_\mathrm{I}$ – a plausible constraint when power control is too complex to implement, such as in decentralized ad-hoc networks. For generality, however, we allow the probe node to employ an arbitrary power $P_0$, not necessarily equal to that of the interfering nodes. We analyze the case of half-duplex transmission, where each device transmits and receives at different time intervals, since full-duplexing capabilities are rare in typical low-cost applications. Nevertheless, the results presented in this paper can be easily modified to account for the full-duplex case.

[3]Whenever possible, we use lowercase letters to denote *deterministic* quantities, and uppercase letters for *stochastic* quantities.

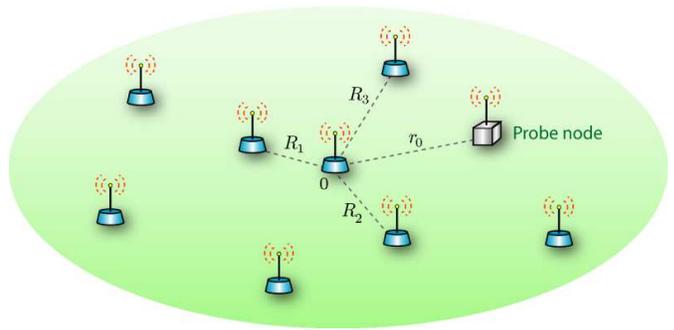

Figure 1. Poisson point process used to model the spatial distribution of nodes. Without loss of generality, we assume the origin of the coordinate system coincides with the probe receiver.

We further consider the scenario where all nodes transmit with the same traffic pattern. In particular, we examine three types of traffic, as depicted in Fig. 2:

1) *Slotted-synchronous traffic:* Similarly to the slotted ALOHA protocol [3], the nodes are synchronized and transmit in slots of duration $L$ seconds.[4] A node transmits in a given slot with probability $q$. The transmissions are independent for different slots and for different nodes.
2) *Slotted-asynchronous traffic:* The nodes transmit in slots of duration $L$ seconds, which are not synchronized with other nodes' time slots. A node transmits in a given slot with probability $q$. The transmissions are independent for different slots or for different nodes.
3) *Exponential-interarrivals traffic:* The nodes transmit packets of duration $L$ seconds. The idle time between packets is exponentially distributed with mean $1/\lambda_\mathrm{p}$.[5]

### C. Wireless Propagation Characteristics

To account for the propagation characteristics of the environment, we consider that the power $P_\mathrm{rx}$ received at a distance $R$ from a transmitter is given by

$$P_\mathrm{rx} = \frac{P_\mathrm{tx} \prod_{k=1}^{K} Z_k}{R^{2b}}, \quad (1)$$

where $P_\mathrm{tx}$ is the average power measured $1$ m away from the transmitter;[6] $b$ is the amplitude loss exponent;[7] and $\{Z_k\}$ are independent random variables (RVs) which account for propagation effects such as multipath fading and shadowing.

[4]By convention, we define these types of traffic with respect to the receiver clock. In the typical case where the propagation delays with respect to the packet length can be ignored, all nodes in the plane observe exactly the same packet arrival process.

[5]This is equivalent to each node using an $M/D/1/1$ queue for packet transmission, characterized by a Poisson arrival process with rate $\lambda_\mathrm{p}$, constant service time $L$, single server, and maximum capacity of one packet.

[6]Unless otherwise stated, we will simply refer to $P_\mathrm{tx}$ as the "transmit power".

[7]Note that the *amplitude loss exponent* is $b$, while the corresponding *power loss exponent* is $2b$.

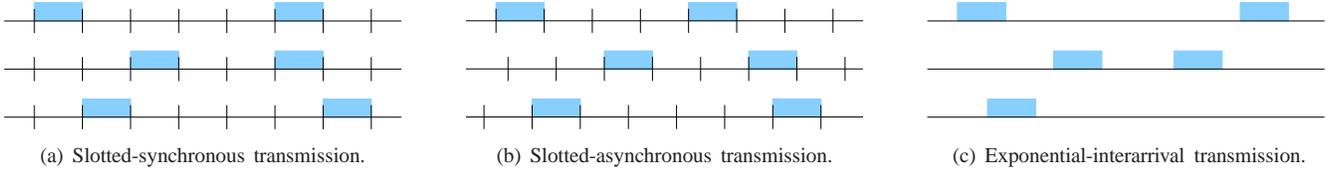

Figure 2. Three types of packet traffic, as observed by the node at the origin.

The term $1/R^{2b}$ accounts for the path loss with distance $R$, where the amplitude loss exponent $b$ is environment-dependent and can approximately range from $0.8$ (e.g., hallways inside buildings) to $4$ (e.g., dense urban environments), with $b = 1$ corresponding to free space propagation [29]. This paper carries out the analysis generally in terms of $\{Z_k\}$, and therefore our results are valid for *any* wireless propagation effect. For illustration purposes, we consistently analyze four typical propagation scenarios throughout this paper:

1) *Path loss only:* $K = 1$ and $Z_1 = 1$.
2) *Path loss and log-normal shadowing:* $K = 1$ and $Z_1 = e^{2\sigma G}$, where $G \sim \mathcal{N}(0,1)$.[8] The term $e^{2\sigma G}$ has a log-normal distribution, where $\sigma$ is the shadowing coefficient.[9]
3) *Path loss and Nakagami-m fading:* $K = 1$ and $Z_1 = \alpha^2$, where $\alpha^2 \sim \mathcal{G}(m, \frac{1}{m})$.[10]
4) *Path loss, log-normal shadowing, and Nakagami-m fading:* $K = 2$, $Z_2 = e^{2\sigma G}$ with $G \sim \mathcal{N}(0,1)$, and $Z_1 = \alpha^2$ with $\alpha^2 \sim \mathcal{G}(m, \frac{1}{m})$.

We emphasize that the proposed framework encompasses a wider variety of propagation effects other than these four cases, such as Rician fading.

## III. CONNECTIVITY-BASED ANALYSIS

In this section, we analyze the throughput of the probe link from a connectivity perspective. In such approach, a node can only hear the transmissions from a finite number of nodes (called *audible interferers*), whose received power exceeds some threshold. For a node to successfully receive the desired packet, it must not collide with any other packet from the audible interferers.[11] Therefore, we start with the statistical characterization of the number of audible nodes, and then use the result to analyze the throughput.

### A. Distribution of the Number of Audible Nodes

We start by defining the concept of audible node.

---
[8]We use $\mathcal{N}(\mu, \sigma^2)$ to denote a Gaussian distribution with mean $\mu$ and variance $\sigma^2$.

[9]This model for combined path loss and log-normal shadowing can be expressed in logarithmic form [29]–[31], such that the channel loss in dB is given by $L_{\text{dB}} = k_0 + k_1 \log_{10} r + \sigma_{\text{dB}} G$, where $G \sim \mathcal{N}(0,1)$. The environment-dependent parameters $(k_0, k_1, \sigma_{\text{dB}})$ can be related to $(b, \sigma)$ as follows: $k_0 = 0$, $k_1 = 20b$ and $\sigma_{\text{dB}} = \frac{20}{\ln 10} \sigma$. The parameter $\sigma_{\text{dB}}$ is the standard deviation of the channel loss in dB (or, equivalently, of the received SNR in dB), and typically ranges from 6 to 12.

[10]We use $\mathcal{G}(x, \theta)$ to denote a gamma distribution with mean $x\theta$ and variance $x\theta^2$.

[11]Our connectivity-based analysis represents a generalization of the *collision model* considered in the context of legacy ALOHA systems [3], in the sense that it encompasses the spatial distribution of nodes, arbitrary propagation effects, and arbitrary packet traffic.

---

*Definition 3.1 (Audible Node):* A node is *audible* to another node if the corresponding received power in (1) satisfies $P_{\text{rx}} \geq P^*$, where $P^*$ denotes some threshold (e.g., related to the sensitivity of the receiver). Otherwise, the node is said to be *inaudible*.

Note that this definition of audible node is a generalization of the notion in [12], which is limited to the case of path loss and log-normal shadowing. Our goal is to statistically characterize the number $N_{\text{A}}$ of nodes that are audible to a given node, as depicted in Fig. 3. From (1), a node $i$ is audible if it satisfies

$$P_{\text{rx},i} = P_{\text{I}} \prod_{k=1}^{K} Z_{i,k} / R_i^{2b} \geq P^*, \quad (2)$$

where $P_{\text{I}}$ is the power transmitted by each interferer; the sequence $\{Z_{i,k}\}$ denotes the propagation effects associated with node $i$, assumed IID in $i$; and $R_i$ denotes the distance from node $i$ to the origin. Thus, we conclude that $N_{\text{A}}$ is a RV, since it is determined by the the random propagation effects and random node positions. To derive the distribution of $N_{\text{A}}$, we mark each node $i$ as either *audible* if condition (2) is satisfied, or *inaudible* otherwise. Note that the markings depend on the position of the marked nodes, but are otherwise independent of each other, since the propagation effects $\{Z_{i,k}\}$ are also independent for different nodes $i$. We can therefore apply the marking theorem [32, Section 5.2], and conclude that the random set of audible nodes forms a two-dimensional (but non-homogeneous) Poisson process. An immediate consequence of this is that the total number $N_{\text{A}}$ of audible nodes is a discrete Poisson RV, i.e.,[12]

$$N_{\text{A}} \sim \mathcal{P}(\mu_{\text{A}}),$$

where the mean $\mu_{\text{A}}$ can be written using [32, Eq. (5.9)] as

$$\begin{aligned}
\mu_{\text{A}} &= \int\!\!\int_{\mathbb{R}^2} p(\mathbf{x}, \text{audible}) \lambda d\mathbf{x} \\
&= \mathbb{E}_{\{Z_k\}} \left\{ \int\!\!\int_{\mathbb{R}^2} p(\mathbf{x}, \text{audible}|\{Z_k\}) \lambda d\mathbf{x} \right\} \\
&= \mathbb{E}_{\{Z_k\}} \{\mu_{\text{A}|\{Z_k\}}\} \\
&= \mathbb{E}_{\{Z_k\}} \left\{ \pi \lambda \left( \frac{P_{\text{I}} \prod_{k=1}^{K} Z_k}{P^*} \right)^{1/b} \right\} \\
&= \pi \lambda \left( \frac{P_{\text{I}}}{P^*} \right)^{1/b} \prod_{k=1}^{K} \mathbb{E}\{Z_k^{1/b}\}. \quad (3)
\end{aligned}$$

In this derivation, we used the fact that when conditioned on the random propagation effects $\{Z_k\}$, a node is audible if it is inside a circle of radius $r_d = (P_I \prod_{k=1}^K Z_k/P^*)^{1/2b}$. The resulting $\mu_{A|\{Z_k\}}$ is simply the area $\pi r_d^2$ of such circle, multiplied by spatial density $\lambda$ of nodes, leading to (3). The impact of the various propagation effects is now evident: each propagation effect $Z_k$ simply contributes with its own multiplicative term in (3). Specifically, the average number $\mu_A$ of audible nodes depends only on the $1/b$-order moments of the random propagation effects $\{Z_k\}$, and not on their entire probability distributions.

---
[12]We use $\mathcal{P}(\mu)$ to denote a discrete Poisson distribution with mean $\mu$.

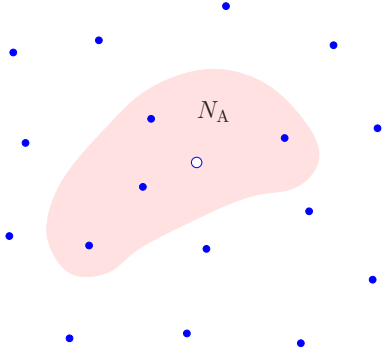

Figure 3. Audible nodes to another node. In this example, the white node has $N_A = 4$ audible nodes.

### B. Effect of the Propagation Characteristics on $\mu_A$

The result in (3) accounts for a wide range of propagation conditions, including the four scenarios analyzed in what follows.

*1) Path loss only:* In this case, $Z_1 = 1$ and (3) reduces to

$$\mu_A = \pi \lambda \left(\frac{P_I}{P^*}\right)^{1/b}. \tag{4}$$

Note that this special case was also obtained in [12], [16].

*2) Path loss and log-normal shadowing:* In this case, $Z_1 = e^{2\sigma G}$ with $G \sim \mathcal{N}(0, 1)$. Using the moment property of log-normal RVs, we have that $\mathbb{E}\{e^{xG}\} = e^{x^2/2}$, $x \geq 0$, and thus (3) reduces to

$$\mu_A = \pi \lambda \left(\frac{P_I}{P^*}\right)^{1/b} \exp\left(\frac{2\sigma^2}{b^2}\right). \tag{5}$$

Note that this particular equation was also obtained in [12] for the case of path loss and log-normal shadowing only, solving a differential equation for the MGF of $N_A$.

*3) Path loss and Nakagami-$m$ fading:* In this case, $Z_1 = \alpha^2$ with $\alpha^2 \sim \mathcal{G}(m, \frac{1}{m})$. Using the moment relation for gamma RVs [33], we have that $\mathbb{E}\{(\alpha^2)^x\} = \frac{\Gamma(m+x)}{m^x \Gamma(m)}$, $x \geq 0$, where $\Gamma(x) = \int_0^\infty t^{x-1} e^{-t} dt$ denotes the gamma function. Thus, (3) reduces to

$$\mu_A = \pi \lambda \left(\frac{P_I}{P^*}\right)^{1/b} \frac{\Gamma\left(m + \frac{1}{b}\right)}{m^{1/b} \Gamma(m)}. \tag{6}$$

For the particular case of Rayleigh fading ($m = 1$), this simplifies to

$$\mu_A = \pi \lambda \left(\frac{P_I}{P^*}\right)^{1/b} \Gamma\left(1 + \frac{1}{b}\right).$$

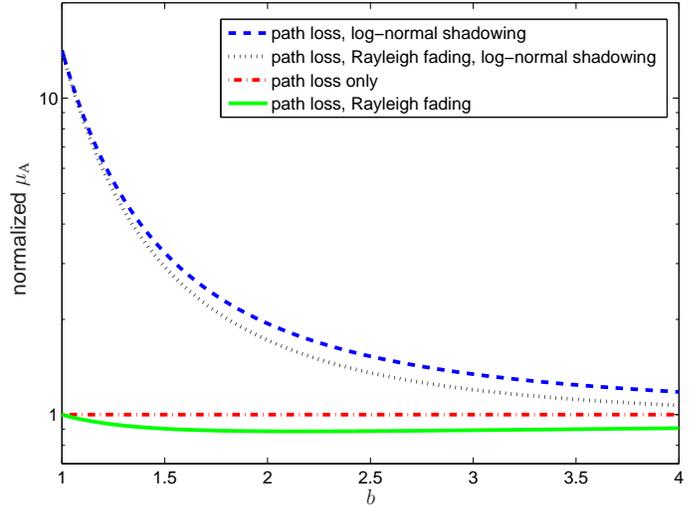

Figure 4. Normalized mean number of audible nodes, $\frac{\mu_A}{\pi \lambda (P_I/P^*)^{1/b}}$, versus the amplitude loss exponent $b$, for various wireless propagation characteristics ($\sigma_{dB} = 10$).

*4) Path loss, log-normal shadowing, and Nakagami-$m$ fading:* Combining scenarios 2 and 3, we have that

$$\mu_A = \pi \lambda \left(\frac{P_I}{P^*}\right)^{1/b} \frac{\Gamma\left(m + \frac{1}{b}\right)}{m^{1/b} \Gamma(m)} \exp\left(\frac{2\sigma^2}{b^2}\right). \tag{7}$$

For the particular case of Rayleigh fading ($m = 1$), this simplifies to

$$\mu_A = \pi \lambda \left(\frac{P_I}{P^*}\right)^{1/b} \Gamma\left(1 + \frac{1}{b}\right) \exp\left(\frac{2\sigma^2}{b^2}\right).$$

From this result we can obtain the probability of node isolation $\mathbb{P}\{N_A = 0\} = e^{-\mu_A}$, which was also derived in [11] using a more involved argument, in the special case of path loss, Rayleigh fading ($m = 1$), and log-normal shadowing only.

*5) Discussion:* The value of $\mu_A$ in these four propagation scenarios is compared in Fig. 4. Considering that $b > 1$, $m \geq \frac{1}{2}$, and $\sigma > 0$, we can show that $\frac{\Gamma(m+\frac{1}{b})}{m^{1/b} \Gamma(m)} < 1 < \exp\left(\frac{2\sigma^2}{b^2}\right)$. Therefore, the log-normal shadowing always improves the connectivity of the network, leading to a larger $\mu_A$ than in the "path loss only" case. The Nakagami-$m$ fading, on the other hand, always worsens the connectivity of the network by decreasing $\mu_A$, in comparison to the "path loss only" case. A similar analysis can be performed for any wireless propagation effect with an arbitrary distribution, or any combination of different propagation effects.

### C. Probe Link Throughput

We now use the results developed in Sections III-A and III-B to determine the throughput of the probe link, subject to the possible packet collisions with other interfering nodes. We start by defining the concept of connectivity-based throughput.

*Definition 3.2 (Connectivity-based Throughput):* The connectivity-based throughput $\mathcal{T}$ of a link is the probability that a packet is successfully received during an interval equal to the packet duration $L$. For a packet to be received received, it cannot collide with other packets from audible transmitters.

Using the definition above, we can write the throughput $\mathcal{T}$ as

$$\mathcal{T} = \mathbb{P}\{\text{probe transmits}\}\mathbb{P}\{\text{receiver silent}\} \\ \times \mathbb{P}\{\text{probe audible}\}\mathbb{P}\{\text{no collision with audible nodes}\}. \tag{8}$$

The first probability term, which we denote by $p_T$, depends on the type of packet traffic. The second term, which we denote by $p_S$, also depends on the type of packet traffic and corresponds to the probability that the node at the origin is silent (i.e., does not transmit) during the transmission of the packet by the probe node. This second term is necessary because the nodes are half-duplex, so they cannot transmit and receive simultaneously. We denote the third term by $p_A$, which corresponds to the probability that the probe node is audible to the node located at the origin. The value of $p_A$ depends on the wireless propagation characteristics, and can be expressed as

$$p_A = \mathbb{P}_{\{Z_{0,k}\}} \left\{ \frac{P_0 \prod_{k=1}^{K} Z_{0,k}}{r_0^{2b}} \geq P^* \right\}, \tag{9}$$

where the subscript 0 refers to the probe link. Lastly, Appendix A shows that fourth term can be written as

$$\mathbb{P}\{\text{no collision with audible nodes}\} = \exp\left(-\mu_A(1-p_S)\right), \tag{10}$$

where $\mu_A$ is generally given in (3). We can then rewrite (8) as

$$\mathcal{T} = p_T p_S p_A \exp\left(-\mu_A(1-p_S)\right). \tag{11}$$

This expression is general and valid for a variety of propagation conditions as well as traffic patterns. As we will see in the next sections, the propagation characteristics determine $p_A$ and $\mu_A$, while the traffic pattern determines $p_T$ and $p_S$. Furthermore, (11) relies on the expressions for $\mu_A$ developed earlier in this paper.

### D. Effect of the Traffic Pattern on $\mathcal{T}$

We now investigate the effect of three different types of traffic pattern described in Section II-B on the throughput. The traffic pattern affects the throughput $\mathcal{T}$ only through $p_T$ and $p_S$ in (11).

*1) Slotted-synchronous traffic:* In this case, the probability that the probe transmits is $p_T = q$, and the probability that the node at the origin is silent during such transmission is $p_S = 1 - q$.

*2) Slotted-asynchronous traffic:* In this case, the probability that the probe transmits is $p_T = q$, and the probability that the node at the origin is silent during such transmission (i.e., during two adjacent time slots) is $p_S = (1-q)^2$.

*3) Exponential-interarrivals traffic:* In this case, we show in Appendix B that the probability that the probe transmits is $p_T = \frac{\lambda_p L}{1+\lambda_p L}$, and the probability that the node at the origin is silent during such transmission is $p_S = \frac{e^{-\lambda_p L}}{1+\lambda_p L}$.

### E. Effect of the Propagation Characteristics on $\mathcal{T}$

We now determine the effect of four propagation scenarios described in Section II-C on the throughput. Recall that the propagation characteristics affect the throughput $\mathcal{T}$ only through $p_A$ and $\mu_A$ in (11). Thus, we now derive $p_A$ for these specific scenarios, since the corresponding expressions for $\mu_A$ have already been determined in Section III-B.

*1) Path loss only :* In this case, the probe node is audible if it is located in a circle of radius $(P_0/P^*)^{1/2b}$ around the origin. Since $r_0$ is deterministic, we have

$$p_A = \begin{cases} 1, & r_0 \leq \left(\frac{P_0}{P^*}\right)^{1/2b}, \\ 0, & \text{otherwise.} \end{cases}$$

*2) Path loss and log-normal shadowing:* In this case, (9) reduces to $p_A = \mathbb{P}_{G_0}\{e^{2\sigma G_0} \geq P^* r_0^{2b}/P_0\}$, where $G_0 \sim \mathcal{N}(0,1)$. Using the Gaussian $Q$-function, we obtain

$$p_A = Q\left(\frac{1}{2\sigma} \ln\left(\frac{P^* r_0^{2b}}{P_0}\right)\right).$$

*3) Path loss and Nakagami-$m$ fading:* In this case, (9) reduces to $p_A = \mathbb{P}_{\alpha_0}\{\alpha_0^2 \geq P^* r_0^{2b}/P_0\}$, where $\alpha_0^2 \sim \mathcal{G}(m, \frac{1}{m})$. Using the cumulative distribution function (CDF) of a gamma RV, we obtain

$$p_A = 1 - \frac{1}{\Gamma(m)} \gamma_{\text{inc}}\left(m, \frac{P^* r_0^{2b} m}{P_0}\right), \tag{12}$$

where $\gamma_{\text{inc}}(a,x) = \int_0^x t^{a-1} e^{-t} dt$ is the lower incomplete gamma function. For integer $m$, we can express $\gamma_{\text{inc}}(a,x)$ in closed form [34], so that (12) simplifies to

$$p_A = 1 - \frac{(m-1)!}{\Gamma(m)}\left(1 - \sum_{k=0}^{m-1} \frac{\nu_1^k e^{-\nu_1}}{k!}\right),$$

where

$$\nu_1 = \frac{P^* r_0^{2b} m}{P_0}.$$

For the particular case of Rayleigh fading ($m=1$), we obtain

$$p_A = \exp\left(-\frac{P^* r_0^{2b}}{P_0}\right).$$

*4) Path loss, log-normal shadowing, and Nakagami-$m$ fading:* In this case, (9) reduces to $p_A = \mathbb{P}_{\alpha_0, G_0}\{\alpha_0^2 e^{2\sigma G_0} \geq P^* r_0^{2b}/P_0\}$. Conditioning on $G_0$, using the CDF of a gamma RV, and then averaging over $G_0$, we obtain

$$p_A = 1 - \frac{1}{\Gamma(m)} \mathbb{E}_{G_0}\left\{\gamma_{\text{inc}}\left(m, \frac{P^* r_0^{2b} m}{P_0 e^{2\sigma G_0}}\right)\right\}. \tag{13}$$

Appendix C shows that if we consider $m$ to be integer and approximate the MGF of the log-normal RV $e^{-2\sigma G_0}$ by a Gauss-Hermite series, $p_A$ simplifies to

$$p_A \approx 1 - \frac{(m-1)!}{\Gamma(m)}\left(1 - \frac{1}{\sqrt{\pi}} \sum_{k=0}^{m-1} \sum_{n=1}^{N_p} H_{x_n} \frac{\nu_2^k e^{-\nu_2}}{k!}\right), \tag{14}$$

where

$$\nu_2 = \frac{P^* r_0^{2b} m}{P_0} e^{2\sqrt{2}\sigma x_n};$$

and $x_n$ and $H_{x_n}$ are, respectively, the zeros and the weights of the $N_p$-order Hermite polynomial. Both $x_n$ and $H_{x_n}$ are tabulated in [34, Table 25.10] for various polynomial orders $N_p$. Typically, setting $N_p = 12$ ensures that the approximation in (14) is extremely accurate [35]. For the particular case of Rayleigh fading ($m = 1$), (14) simplifies to

$$p_A \approx \frac{1}{\sqrt{\pi}} \sum_{n=1}^{N_p} H_{x_n} \exp\left(-\frac{P^* r_0^{2b}}{P_0} e^{2\sqrt{2}\sigma x_n}\right).$$

### F. Discussion

Using the results derived in this section, we can obtain insights into the behaviour of the throughput as a function of important network parameters, such as the type of propagation characteristics or traffic pattern. In particular, the throughput in the slotted-synchronous and slotted-asynchronous cases can be related as follows. Considering that $\mu_A > 0$, we can easily show that $q(1-q)e^{-\mu_A q} \geq q(1-q)^2 e^{-\mu_A [1-(1-q)^2]}$, with equality iff $q = 0$ or $q = 1$. Thus, the throughput of a wireless network is higher for slotted-synchronous traffic than for slotted-asynchronous traffic, regardless of the specific propagation conditions.[13] We will illustrate this property in Section V using numerical examples. The reason for the higher throughput performance in the synchronous case is that a packet can potentially collide with only *one packet* transmitted by another node, while in the asynchronous case it can collide with any of the *two packets* in adjacent time slots (see Fig. 2).

We can also analyze how the throughput of the probe link depends on the interfering nodes, which are characterized by their spatial density $\lambda$ and the transmit power $P_I$. Specifically, the throughput $\mathcal{T}$ in the general expression (11) depends on $\lambda$ and $P_I$ only through $\mu_A$. From (3), we have that $\mu_A \propto \lambda P_I^{1/b}$, and thus $\mathcal{T} \propto \exp(-c\lambda P_I^{1/b})$, for $b > 1$ and some proportionality constant $c$.[14] This quantifies exactly how the throughput of the probe link degrades with an increase in the density and power of the interferers; in particular, it shows that the *throughput is more sensitive to an increase in the spatial density of the interferers, than to an increase in their power*. This observation is valid for any wireless propagation characteristics and traffic pattern.

## IV. SINR-BASED ANALYSIS

In this section, we analyze the throughput of the probe link from a SINR perspective. In such approach, a node can hear the transmissions from *all* the nodes in the two-dimensional plane [1], [22], [36], not just from a finite number of audible nodes, as in the connectivity-based approach developed in Section III. For a node to successfully receive a desired packet, the SINR must exceed some threshold.[15] Therefore, we start with the statistical characterization of the SINR, and then use the results to analyze the throughput.

---

[13] Note that $p_A$ and $\mu_A$ in (11) do not depend on the traffic type.
[14] We use $\propto$ to denote proportionality.
[15] Our SINR-based analysis represents a generalization of the *capture model* considered in the context of ALOHA systems [4]–[7] and of the *physical model* considered in [22], in the sense that it encompasses the spatial distribution of nodes, arbitrary propagation effects, and arbitrary packet traffic.

### A. Signal-to-Interference-Plus-Noise Ratio

Typically, the distances $\{R_i\}$ and propagation effects $\{Z_{i,k}\}$ are slowly-varying and remain approximately constant during the packet duration $L$. In this quasi-static scenario, it is insightful to define the SINR conditioned on a given realization of those RVs. As we shall see, this naturally leads to the derivation of an *SINR outage probability*, which in turn determines the throughput. We start by formally defining the concept of SINR.

*Definition 4.1 (Signal-to-Interference-Plus-Noise Ratio):* The signal-to-interference-plus-noise ratio associated with the node at the origin is defined as

$$\mathsf{SINR} \triangleq \frac{S}{I+N}, \quad (15)$$

where $S$ is the power of the desired signal received from the probe node, $I$ is the aggregate interference power received from all other nodes in the network, and $N$ is the (constant) noise power. Both $S$ and $I$ depend on a given realization of $\{R_i\}$, $i = 1\ldots\infty$, and $\{Z_{i,k}\}$, $i = 0\ldots\infty$, $k = 1\ldots K$.

Using (1), the desired signal power $S$ can be written as

$$S = \frac{P_0 \prod_{k=1}^{K} Z_{0,k}}{r_0^{2b}}. \quad (16)$$

Similarly, the aggregate interference power $I$ can be written as

$$I = \sum_{i=1}^{\infty} \frac{P_I \Delta_i \prod_{k=1}^{K} Z_{i,k}}{R_i^{2b}}, \quad (17)$$

where $P_I$ is the transmit power associated with each interferer, and $\Delta_i \in [0,1]$ is the (random) duty-cycle factor associated with interferer $i$. As we shall see, the RV $\Delta_i$ accounts for the different traffic patterns of nodes, and is equal to the fraction of the packet duration $L$ during which interferer $i$ is effectively transmitting. Note that since $S$ and $I$ depend on the random node positions and random propagation effects, they can be seen as RVs whose value is different for each realization of those random quantities. Furthermore, we showed in [1] that the RV $I$ has a *skewed stable distribution* [37] given by[16]

$$I \sim \mathcal{S}\bigg(\alpha = \frac{1}{b}, \beta = 1, \quad (18)$$

$$\gamma = \pi\lambda C_{1/b}^{-1} P_I^{1/b} \mathbb{E}\{\Delta_i^{1/b}\} \prod_{k=1}^{K} \mathbb{E}\{Z_{i,k}^{1/b}\}\bigg)$$

where $b > 1$, and $C_x$ is defined as

$$C_x \triangleq \begin{cases} \frac{1-x}{\Gamma(2-x)\cos(\pi x/2)}, & x \neq 1, \\ \frac{2}{\pi}, & x = 1. \end{cases} \quad (19)$$

As we shall see, the probe link throughput depends on the traffic pattern of the nodes only through $\mathbb{E}\{\Delta_i^{1/b}\}$ in (18).

---

[16] We use $\mathcal{S}(\alpha, \beta, \gamma)$ to denote a stable distribution with characteristic exponent $\alpha \in (0,2]$, skewness $\beta \in [-1,1]$, and dispersion $\gamma \in [0,\infty)$. The corresponding characteristic function is [37]

$$\phi(w) = \begin{cases} \exp\left[-\gamma|w|^\alpha \left(1 - j\beta\,\mathrm{sign}(w)\tan\frac{\pi\alpha}{2}\right)\right], & \alpha \neq 1, \\ \exp\left[-\gamma|w|\left(1 + j\frac{2}{\pi}\beta\,\mathrm{sign}(w)\ln|w|\right)\right], & \alpha = 1. \end{cases}$$

## B. Probe Link Throughput

We now use the results developed in Section IV-A to characterize the throughput of the probe link, subject to the aggregate network interference. We start by defining the concept of SINR-based throughput.

*Definition 4.2 (SINR-based Throughput):* The *SINR-based throughput* $\mathcal{T}$ of a link is the probability that a packet is successfully received during an interval equal to the packet duration $L$. For a packet to be successfully received, the SINR of the link must exceed some threshold.

Using the definition above, we can write the throughput $\mathcal{T}$ as

$$\mathcal{T} = \mathbb{P}\{\text{probe transmits}\}\mathbb{P}\{\text{receiver silent}\}\mathbb{P}\{\text{no outage}\}. \quad (20)$$

The first and second probability terms were defined in Section III-C as $p_\text{T}$ and $p_\text{S}$, respectively. The third term is simply $\mathbb{P}\{\text{SINR} \geq \theta^*\}$, where $\theta^*$ is a predetermined threshold that ensures reliable packet communication over the probe link. Using (15), (16), and the law of total probability with respect to RVs $\{Z_{0,k}\}$ and $I$, we can express such probability as

$$\mathbb{P}\{\text{SINR} \geq \theta^*\} = \mathbb{E}_I\left\{\mathbb{P}_{\{Z_{0,k}\}}\left\{\prod_{k=1}^{K} Z_{0,k} \geq \frac{r_0^{2b}\theta^*}{P_0}(I+N)\bigg|I\right\}\right\} \quad (21)$$

or, alternatively, as

$$\mathbb{P}\{\text{SINR} \geq \theta^*\} = \mathbb{E}_{\{Z_{0,k}\}}\left\{F_I\left(\frac{P_0 \prod_{k=1}^{K} Z_{0,k}}{r_0^{2b}\theta^*} - N\right)\right\}, \quad (22)$$

where $F_I(\cdot)$ is the CDF of the stable RV $I$, whose distribution is given in (18). As we shall see, both forms are useful depending on the considered propagation characteristics. We can therefore write the throughput of a wireless network as

$$\mathcal{T} = p_\text{T} p_\text{S} \mathbb{P}\{\text{SINR} \geq \theta^*\}. \quad (23)$$

This expression is general and valid for a variety of propagation conditions as well as traffic patterns. As we will see in the next sections, the propagation characteristics determine only $\mathbb{P}\{\text{SINR} \geq \theta^*\}$, while the traffic pattern determines $p_\text{T}$, $p_\text{S}$, and $\mathbb{P}\{\text{SINR} \geq \theta^*\}$.

## C. Effect of the Traffic Pattern on $\mathcal{T}$

We now investigate the effect of three different types of traffic pattern described in Section II-B on the throughput. Recall that the traffic pattern affects the throughput $\mathcal{T}$ through $p_\text{T}$, $p_\text{S}$, and $\mathbb{P}\{\text{SINR} \geq \theta^*\}$ in (23). The type of packet traffic determines the statistics of the duty-cycle factor $\Delta_i$, and in particular $\mathbb{E}\{\Delta_i^{1/b}\}$ in (18), which in turn affects $\mathbb{P}\{\text{SINR} \geq \theta^*\}$.

*1) Slotted-synchronous traffic:* In this case, $p_\text{T} = q$ and $p_\text{S} = 1 - q$. The duty-cycle factor $\Delta_i$ is a binary RV taking the value $0$ if the interferer $i$ is silent in the considered time slot (with probability $1-q$), and the value $1$ if the interferer $i$ transmits in the slot (with probability $q$). Then, the PDF of $\Delta_i$ can be written as $f_\Delta(x) = (1-q)\delta(x) + q\delta(x-1)$, and after some algebra $\mathbb{E}\{\Delta_i^{1/b}\}$ can be shown to be

$$\mathbb{E}\{\Delta_i^{1/b}\} = q. \quad (24)$$

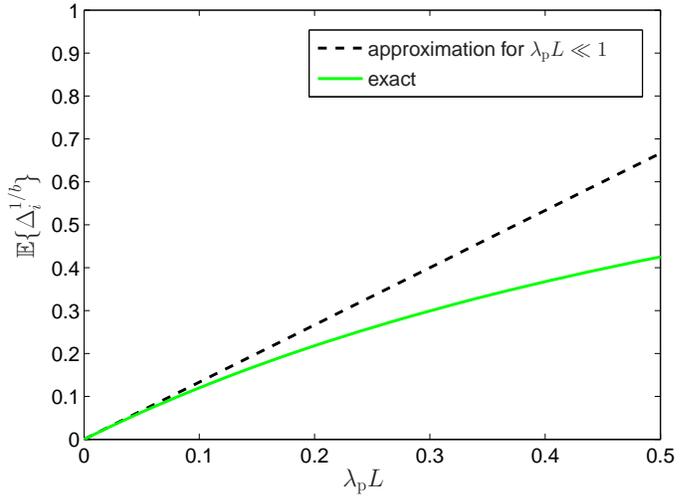

Figure 5. Comparison between aproximated and exact expressions for $\mathbb{E}\{\Delta_i^{1/b}\}$ in the exponential-interarrivals scenario ($b=2$).

*2) Slotted-asynchronous traffic:* In this case, $p_\text{T} = q$ and $p_\text{S} = (1-q)^2$. Using the law of total probability, the duty-cycle factor $\Delta_i$ is $0$ if a node is silent in two adjacent time slots (with probability $(1-q)^2$), a uniform continuous RV in the interval $[0,1]$ if the node transmits in either one of the adjacent time slots, or $1$ if the node transmits in the two adjacent time slots (with probability $q^2$). Then, $f_\Delta(x) = (1-q)^2\delta(x) + 2q(1-q) + q^2\delta(x-1)$ for $0 \leq x \leq 1$, and after some algebra $\mathbb{E}\{\Delta_i^{1/b}\}$ can be shown to be

$$\mathbb{E}\{\Delta_i^{1/b}\} = q^2 + 2q(1-q)\frac{b}{b+1}. \quad (25)$$

*3) Exponential-interarrivals traffic:* In this case, we show in Appendix B that $p_\text{T} = \frac{\lambda_\text{p}L}{1+\lambda_\text{p}L}$, $p_\text{S} = \frac{e^{-\lambda_\text{p}L}}{1+\lambda_\text{p}L}$, and

$$\mathbb{E}\{\Delta_i^{1/b}\} = \frac{2\lambda_\text{p}L}{1+\lambda_\text{p}L} \cdot \mathcal{I}\left(\frac{1}{b}, \lambda_\text{p}L\right) \quad (26)$$
$$+ \frac{(\lambda_\text{p}L)^2}{1+\lambda_\text{p}L} \cdot \mathcal{I}\left(\frac{1}{b}+1, \lambda_\text{p}L\right),$$

where $\mathcal{I}(x,y) = \int_0^1 (1-t)^x e^{-yt} dt$. In the regime where $\lambda_\text{p}L \ll 1$, (26) can be approximated by

$$\mathbb{E}\{\Delta_i^{1/b}\} \approx 2\lambda_\text{p}L\frac{b}{b+1}.$$

The accuracy of this approximation is illustrated in Figure 5. For example, consider a sensor network where each node generates measurements at an average rate of $\lambda_\text{p} = 1\,\text{s}^{-1}$, and encodes them in 128-bit packets, which are then transmitted at 10 Kbit/s. Then, $\lambda_\text{p}L = 0.0128 \ll 1$, and the approximate value for $\mathbb{E}\{\Delta_i^{1/b}\}$ is within 2% of the exact value.

As a remark, note that the result in (24) for the synchronous case can also be derived using the splitting property of Poisson processes [38]. Specifically, since a node transmits

$$\mathbb{P}\{\mathsf{SINR} \geq \theta^*\} = 1 - \frac{(m-1)!}{\Gamma(m)} \left(1 - \sum_{k=0}^{m-1} \sum_{j=0}^{k} \frac{(-\nu_3)^j}{j!} \frac{(\nu_3 N)^{k-j} e^{-\nu_3 N}}{(k-j)!} \left.\frac{d^j \phi_I(s)}{ds^j}\right|_{s=\nu_3}\right) \qquad (30)$$

$$\mathbb{P}\{\mathsf{SINR} \geq \theta^*\} = \exp\left(-\frac{r_0^{2b}\theta^* N}{P_0}\right) \exp\left(-\frac{\pi\lambda C_{1/b}^{-1}\Gamma\left(1+\frac{1}{b}\right)\mathbb{E}\{\Delta_i^{1/b}\}}{\cos\left(\frac{\pi}{2b}\right)}\left(\frac{P_I r_0^{2b}\theta^*}{P_0}\right)^{1/b}\right) \qquad (32)$$

---

in a given slot with probability $q$, the effective spatial density of nodes that contribute to the interference is $\lambda_{\text{eff}} = \lambda q$, and $\gamma$ in (18) reduces to $\gamma = \pi\lambda_{\text{eff}} C_{1/b}^{-1} P_I^{1/b} \prod_{k=1}^{K} \mathbb{E}\{Z_{i,k}^{1/b}\} = \pi\lambda q C_{1/b}^{-1} P_I^{1/b} \prod_{k=1}^{K} \mathbb{E}\{Z_{i,k}^{1/b}\}$. This is precisely the same result obtained by substituting (24) into (18). Although, strictly speaking, the splitting property cannot be used in the case of slotted-asynchronous or exponential-interarrivals traffic, the above discussion suggests that the quantity $\lambda\mathbb{E}\{\Delta_i^{1/b}\}$ can be thought of as an "effective" spatial density of interferers, where $\mathbb{E}\{\Delta_i^{1/b}\} \in [0,1]$ is analogous to a transmission probability.

### D. Effect of the Propagation Characteristics on $\mathcal{T}$

We now determine the effect of four different propagation scenarios described in Section II-C on the throughput. Recall that the propagation characteristics affect the throughput $\mathcal{T}$ only through $\mathbb{P}\{\mathsf{SINR} \geq \theta^*\}$ in (23), and so we now derive such probability for these specific scenarios.

*1) Path loss only:* In this case, the expectation in (22) disappears and we have

$$\mathbb{P}\{\mathsf{SINR} \geq \theta^*\} = F_I\left(\frac{P_0}{r_0^{2b}\theta^*} - N\right), \qquad (27)$$

where the distribution of $I$ in (18) reduces to

$$I \sim \mathcal{S}\left(\alpha = \frac{1}{b},\ \beta = 1,\ \gamma = \pi\lambda C_{1/b}^{-1} P_I^{1/b} \mathbb{E}\{\Delta_i^{1/b}\}\right).$$

Note that the characteristic function of $I$ was also obtained in [26] using the influence function method, for the case of path loss and slotted-synchronous traffic only.

*2) Path loss and log-normal shadowing:* In this case, (21) reduces to $\mathbb{P}\{\mathsf{SINR} \geq \theta^*\} = \mathbb{E}_I\left[\mathbb{P}_{G_0}\{e^{2\sigma G_0} \geq r_0^{2b}\theta^*(I+N)/P_0|I\}\right]$, where $G_0 \sim \mathcal{N}(0,1)$. Using the Gaussian $Q$-function, we obtain

$$\mathbb{P}\{\mathsf{SINR} \geq \theta^*\} = \mathbb{E}_I\left\{Q\left(\frac{1}{2\sigma}\ln\left[\frac{r_0^{2b}\theta^*(I+N)}{P_0}\right]\right)\right\}, \qquad (28)$$

where the distribution of $I$ in (18) reduces to

$$I \sim \mathcal{S}\left(\alpha = \frac{1}{b},\ \beta = 1,\ \gamma = \pi\lambda C_{1/b}^{-1} P_I^{1/b} e^{2\sigma^2/b^2} \mathbb{E}\{\Delta_i^{1/b}\}\right).$$

*3) Path loss and Nakagami-$m$ fading:* In this case, (21) reduces to $\mathbb{P}\{\mathsf{SINR} \geq \theta^*\} =$ $\mathbb{E}_I\left\{\mathbb{P}_{\alpha_0}\left\{\alpha_0^2 \geq \frac{r_0^{2b}\theta^*}{P_0}(I+N)\Big|I\right\}\right\}$, where $\alpha_0^2 \sim \mathcal{G}(m, \frac{1}{m})$. Using the CDF of a gamma RV, we obtain

$$\mathbb{P}\{\mathsf{SINR} \geq \theta^*\} = 1 - \frac{1}{\Gamma(m)}\mathbb{E}_I\left\{\gamma_{\text{inc}}\left(m, \frac{r_0^{2b}\theta^*(I+N)m}{P_0}\right)\right\}, \qquad (29)$$

where the distribution of $I$ in (18) reduces to

$$I \sim \mathcal{S}\left(\alpha = \frac{1}{b},\ \beta = 1,\right.$$
$$\left.\gamma = \pi\lambda C_{1/b}^{-1} P_I^{1/b} \mathbb{E}\{\Delta_i^{1/b}\}\frac{\Gamma\left(m+\frac{1}{b}\right)}{m^{1/b}\Gamma(m)}\right).$$

Appendix D shows that for integer $m$, (29) can be expressed in closed form as (30) given at the top of this page, where

$$\nu_3 = \frac{r_0^{2b}\theta^* m}{P_0},$$

and the moment generating function (MGF) of $I$ is given by

$$\phi_I(s) = \exp\left(-\frac{\pi\lambda C_{1/b}^{-1} P_I^{1/b}\Gamma\left(m+\frac{1}{b}\right)\mathbb{E}\{\Delta_i^{1/b}\}}{m^{1/b}\Gamma(m)\cos\left(\frac{\pi}{2b}\right)} s^{1/b}\right), \qquad (31)$$

for $s \geq 0$. For the particular case of Rayleigh fading ($m=1$), we obtain (32) at the top of this page.

*4) Path loss, log-normal shadowing, and Nakagami-$m$ fading:* In this case, we start with (21), condition on $G_0$, use the result from scenario 3, and then average over $G_0$, leading to

$$\mathbb{P}\{\mathsf{SINR} \geq \theta^*\} \qquad (33)$$
$$= 1 - \frac{1}{\Gamma(m)}\mathbb{E}_{G_0,I}\left\{\gamma_{\text{inc}}\left(m, \frac{r_0^{2b}\theta^*(I+N)m}{P_0 e^{2\sigma G_0}}\right)\right\},$$

where the distribution of $I$ in (18) reduces to

$$I \sim \mathcal{S}\left(\alpha = \frac{1}{b},\ \beta = 1,\right.$$
$$\left.\gamma = \pi\lambda C_{1/b}^{-1} P_I^{1/b} e^{2\sigma^2/b^2}\mathbb{E}\{\Delta_i^{1/b}\}\frac{\Gamma\left(m+\frac{1}{b}\right)}{m^{1/b}\Gamma(m)}\right).$$

For integer $m$, this can be expressed in closed form as given in (34) at the top of the next page, where

$$\nu_4 = \frac{r_0^{2b}\theta^* m}{P_0 e^{2\sigma G_0}},$$

and

$$\phi_I(s) = \exp\left(-\frac{\pi\lambda C_{1/b}^{-1} P_I^{1/b} e^{2\sigma^2/b^2}\Gamma\left(m+\frac{1}{b}\right)\mathbb{E}\{\Delta_i^{1/b}\}}{m^{1/b}\Gamma(m)\cos\left(\frac{\pi}{2b}\right)} s^{1/b}\right)$$

$$\mathbb{P}\{\mathsf{SINR} \geq \theta^*\} = 1 - \frac{(m-1)!}{\Gamma(m)} \left(1 - \sum_{k=0}^{m-1} \sum_{j=0}^{k} \mathbb{E}_{G_0} \left\{ \frac{(-\nu_4)^j}{j!} \frac{(\nu_4 N)^{k-j} e^{-\nu_4 N}}{(k-j)!} \frac{d^j \phi_I(s)}{ds^j}\bigg|_{s=\nu_4} \right\} \right) \quad (34)$$

$$\mathbb{P}\{\mathsf{SINR} \geq \theta^*\} = \mathbb{E}_{G_0} \left\{ \exp\left(-\frac{r_0^{2b}\theta^* N}{P_0 e^{2\sigma G_0}}\right) \exp\left(-\frac{\pi\lambda C_{1/b}^{-1} e^{2\sigma^2/b^2} \Gamma\left(1+\frac{1}{b}\right) \mathbb{E}\{\Delta_i^{1/b}\}}{\cos\left(\frac{\pi}{2b}\right)} \left(\frac{P_\mathrm{I} r_0^{2b}\theta^*}{P_0 e^{2\sigma G_0}}\right)^{1/b}\right) \right\} \quad (35)$$

for $s \geq 0$. The derivation of (34) is completely analogous to that presented in Appendix D, and is omitted here for brevity. For the particular case of Rayleigh fading ($m=1$), we obtain (35) at the top of this page.

*E. Discussion*

Using the results derived in this section, we can obtain insights into the behaviour of the throughput as a function of important network parameters, such as the type of propagation characteristics or traffic pattern. In particular, the throughput in the slotted-synchronous and slotted-asynchronous cases can be related as follows. Considering that $b > 1$, we can easily show that $q \leq q^2 + 2q(1-q)\frac{b}{b+1}$, with equality iff $q=0$ or $q=1$. Therefore, $\mathbb{E}\{\Delta_i^{1/b}\}$ is smaller (or, equivalently, $\mathbb{P}\{\mathsf{SINR} \geq \theta^*\}$ is larger) for the slotted-synchronous case than for the slotted-asynchronous case, regardless of the specific propagation conditions. Furthermore, since $q(1-q) \geq q(1-q)^2$, we conclude that the throughput $\mathcal{T}$ given in (23) is higher for slotted-synchronous traffic than for slotted-asynchronous traffic. We will illustrate this property in Section V using numerical examples. Note that a similar result was obtained in Section III-D for the connectivity-based analysis. Again, the reason for the higher throughput performance in the synchronous case is that a packet can potentially overlap with only *one packet* transmitted by another node, while in the asynchronous case it can collide with any of the *two packets* in adjacent time slots.

We can also analyze how the throughput of the probe link depends on the interfering nodes, which are characterized by their spatial density $\lambda$ and the transmit power $P_\mathrm{I}$. In all the expressions for $\mathbb{P}\{\mathsf{SINR} \geq \theta^*\}$ in Section IV, we can make the parameters $\lambda$ and $P_\mathrm{I}$ appear explicitly by noting that if $I \sim \mathcal{S}(\alpha, \beta, \gamma)$, then $\widetilde{I} = \gamma^{-1/\alpha} I \sim \mathcal{S}(\alpha, \beta, 1)$ is a normalized version of $I$ with unit dispersion. Thus, we can for example rewrite (22) as

$$\mathbb{P}\{\mathsf{SINR} \geq \theta^*\} = \quad (36)$$
$$\mathbb{E}_{\{Z_{0,k}\}} \left\{ F_{\widetilde{I}} \left( \frac{\frac{P_0 \prod_{k=1}^{K} Z_{0,k}}{r_0^{2b}\theta^*} - N}{P_\mathrm{I} \left[\pi\lambda C_{1/b}^{-1} \mathbb{E}\{\Delta_i^{1/b}\} \prod_{k=1}^{K} \mathbb{E}\{Z_{i,k}^{1/b}\}\right]^b} \right) \right\},$$

where $\widetilde{I} \sim \mathcal{S}\left(\alpha = \frac{1}{b}, \beta = 1, \gamma = 1\right)$ only depends on the amplitude loss exponent $b$. Furthermore, since $F_{\widetilde{I}}(\cdot)$ is monotonically increasing with respect to its argument, we conclude that $\mathbb{P}\{\mathsf{SINR} \geq \theta^*\}$ and therefore the throughput $\mathcal{T}$ are monotonically decreasing with $\lambda$ and $P_\mathrm{I}$. In particular, since $b > 1$, the throughput is more sensitive to an increase in the *spatial density* of the interferers, than with an increase in their *transmitter power*. This analysis is valid for any wireless propagation characteristics and traffic pattern.

## V. NUMERICAL RESULTS

Figures 6 and 7 quantify the throughput for the connectivity-based analysis of Section III, showing their dependence on various parameters involved, such as the transmission probability, transmit power, interferer spatial density, type of packet traffic, and wireless propagation effects. In Figure 6, we observe that the throughput is higher for slotted-synchronous traffic than for slotted-asynchronous traffic, as it was shown in Section III-D. Figure 7 shows that in an homogeneous scenario where all nodes transmit with the same power (i.e., $P_0 = P_\mathrm{I} = P$), the throughput $\mathcal{T}$ achieves a maximum for some optimum power level $P_\mathrm{opt}$. We also verify that $\mathcal{T} \to 0$ as $P \to 0$, because the term $p_\mathrm{A} = \mathbb{P}\{\text{probe audible}\}$ in (11) approaches zero. In addition, $\mathcal{T} \to 0$ as $P \to \infty$, since the average number $\mu_\mathrm{A}$ of audible nodes in (11) approaches infinity, thereby increasing the probability of packet collision.

Figures 8 and 9 quantify the throughput for the SINR-based analysis of Section IV. In Figure 8, we observe that, as in the connectivity-based analysis, the throughput is higher for slotted-synchronous traffic than for slotted-asynchronous traffic. This fact was demonstrated in Section IV-C. Figure 9 shows that the throughput $\mathcal{T}$ decreases monotonically with the spatial density $\lambda$ of the interferers, as discussed in Section IV-E.

Figure 10 compares the $\mathcal{T} - \lambda$ curves for both connectivity-based and SINR-based analyses. Note that the two models presented in this paper are defined in terms of distinct sets of parameters. Specifically, the connectivity-based model depends on the audibility threshold $P^*$, while the SINR-based model depends on the SINR threshold $\theta^*$ and the noise power $N$. Each model may be more appropriate depending on the specific application.

## VI. CONCLUSION

In this paper, we introduced a mathematical framework for the characterization of connectivity and throughput in wireless networks. Our work generalizes and unifies various results scattered throughout the literature, by accommodating arbitrary wireless propagation effects, as well as arbitrary traffic patterns. This allows us to draw conclusions about the performance of the network that are more general than the previously available literature.

We proved that the number of audible nodes $N_\mathrm{A}$ in a spatial Poisson process is a discrete Poisson RV, whose mean $\mu_\mathrm{A}$ we characterized for arbitrary propagation scenarios.

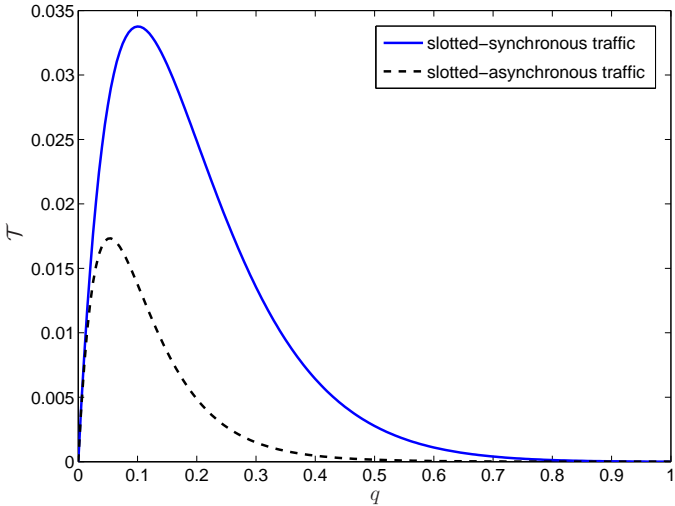

Figure 6. Throughput $\mathcal{T}$ versus the transmission probability $q$, for various types of packet traffic (connectivity-based approach, path loss and Rayleigh fading, $P_0/P^* = P_\mathrm{I}/P^* = 10$, $\lambda = 1\,\mathrm{m}^{-2}$, $b = 2$, $r_0 = 1\,\mathrm{m}$).

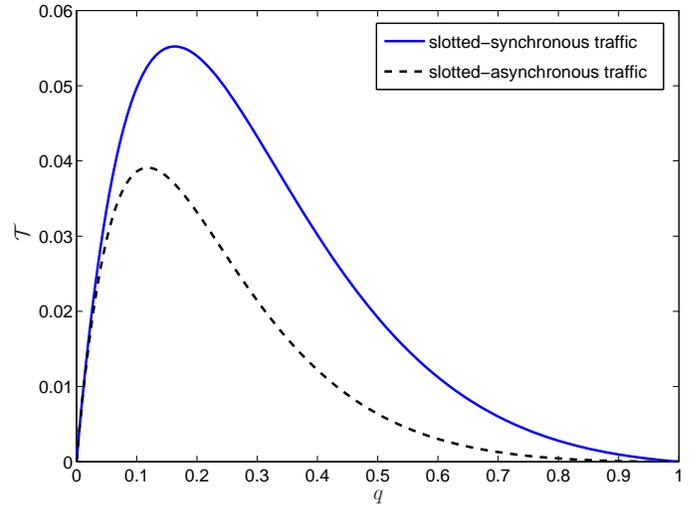

Figure 8. Throughput $\mathcal{T}$ versus the transmission probability $q$, for various types of packet traffic (SINR-based approach, path loss and Rayleigh fading, $P_0/N = P_\mathrm{I}/N = 10$, $\theta^* = 1$, $\lambda = 1\,\mathrm{m}^{-2}$, $b = 2$, $r_0 = 1\,\mathrm{m}$).

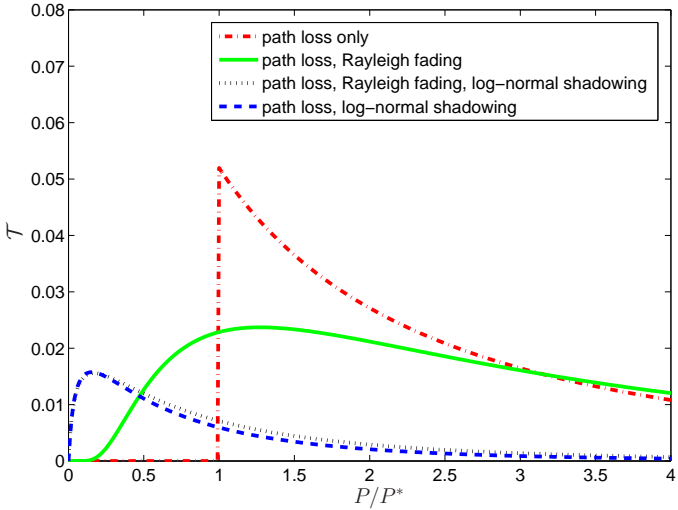

Figure 7. Throughput $\mathcal{T}$ versus $P/P^*$, with $P = P_0 = P_\mathrm{I}$, for various wireless propagation characteristics (connectivity-based approach, slotted-synchronous traffic, $q = 0.5$, $\lambda = 1\,\mathrm{m}^{-2}$, $b = 2$, $r_0 = 1\,\mathrm{m}$, $\sigma_\mathrm{dB} = 10$).

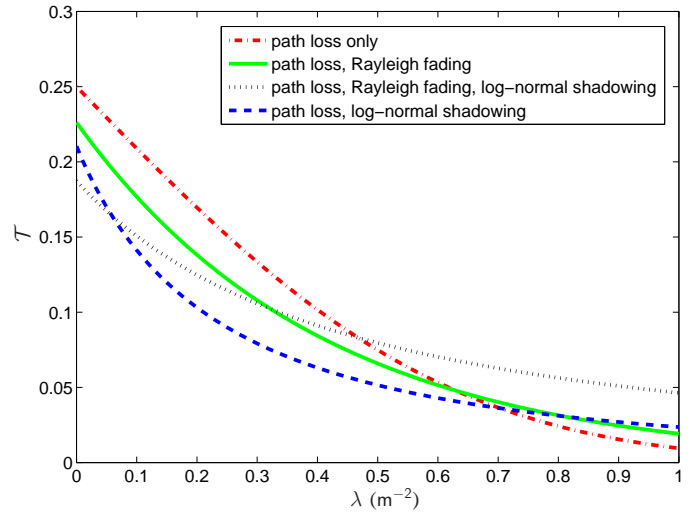

Figure 9. Throughput $\mathcal{T}$ versus the interferer spatial density $\lambda$, for various wireless propagation characteristics (SINR-based approach, slotted-synchronous traffic, $P_0/N = P_\mathrm{I}/N = 10$, $\theta^* = 1$, $q = 0.5$, $b = 2$, $r_0 = 1\,\mathrm{m}$, $\sigma_\mathrm{dB} = 10$).

Specifically, we showed that $\mu_\mathrm{A}$ depends on the $1/b$-order moments of the random propagation effects $\{Z_k\}$, and not on their entire probability distributions. Furthermore, the effect of the propagation characteristics can be decoupled, since each characteristic contributes with its own separate factor to $\mu_\mathrm{A}$. We also showed that log-normal shadowing always improves the connectivity of a wireless network when compared to a scenario with only path loss, while Nakagami-$m$ fading always worsens such connectivity.

We provided expressions for the CDF of the SINR of a link subject to the aggregate interference from all the interferers in the network and noise. We showed how the type of propagation scenario and the packet traffic affect such probabilistic characterization.

We obtained expressions for the link throughput, both from a connectivity perspective and an SINR perspective. We analyzed the effect of the propagation characteristics and the packet traffic on the throughput. Specifically, we showed that the throughput is higher for slotted-synchronous traffic than for slotted-asynchronous traffic, regardless of the specific propagation conditions. We also showed that the throughput degrades faster with an increase in the spatial density of the interferers, than with an increase in their power, regardless of the specific propagation conditions and traffic pattern.

## APPENDIX A
## DERIVATION OF $\mathbb{P}\{\text{no collision}\}$ IN (10)

Using the the law of total probability and the fact that $N_\mathrm{A} \sim \mathcal{P}(\mu_\mathrm{A})$, we can write

$$\mathbb{P}\{\text{no collision}\} = \sum_{n=0}^{\infty} \mathbb{P}\{\text{no collision}|N_\mathrm{A} = n\}\mathbb{P}(N_\mathrm{A} = n)$$

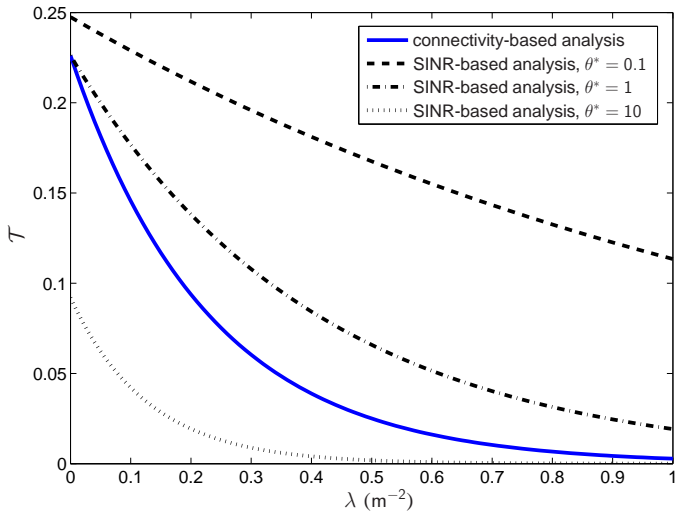

Figure 10. Throughput $\mathcal{T}$ versus the interferer spatial density $\lambda$, for both connectivity-based and SINR-based analyses (path loss and Rayleigh fading, slotted-synchronous traffic, $P^* = N$, $P_0/N = P_\text{I}/N = 10$, $q = 0.5$, $b = 2$, $r_0 = 1\,\text{m}$).

$$= \sum_{n=0}^{\infty} p_\text{S}^n \frac{\mu_\text{A}^n e^{-\mu_\text{A}}}{n!}$$

$$= e^{-\mu_\text{A}} e^{\mu_\text{A} p_\text{S}} \underbrace{\sum_{n=0}^{\infty} \frac{(\mu_\text{A} p_\text{S})^n e^{-\mu_\text{A} p_\text{S}}}{n!}}_{=1}$$

$$= \exp\left(-\mu_\text{A}(1 - p_\text{S})\right).$$

This is the result in (10) and the proof is complete.

## APPENDIX B
## DERIVATION OF $p_\text{T}$, $p_\text{S}$, AND $\mathbb{E}\{\Delta_i^{1/b}\}$ FOR EXPONENTIAL-INTERARRIVALS TRAFFIC

In the case of exponential-interarrivals, each user employs an $M/D/1/1$ queue for packet transmission, characterized by a Poisson arrival process with rate $\lambda_\text{p}$, constant service time $L$, single server, and maximum capacity of one packet. If $Q(t) \in \{0,1\}$ denotes the number of packets in the queue at time $t$, the steady-state probabilities of the queue are $\pi_0 = \mathbb{P}\{Q(t) = 0\} = \frac{1}{1+\lambda_\text{p}L}$, and $\pi_1 = \mathbb{P}\{Q(t) = 1\} = \frac{\lambda_\text{p}L}{1+\lambda_\text{p}L}$ [39]. Then, the parameter $p_\text{T}$ corresponding to the probability of probe transmission is simply given by $p_\text{T} = \pi_1$. To determine $p_\text{S}$, we note that the probe receiver is silent during the probe transmission if the state of its own queue is $Q = 0$ at the beginning of the probe packet, *and* there is no packet arrival in its queue during the following $L$ seconds. As a result, we have $p_\text{S} = \pi_0 e^{-\lambda_\text{p}L}$.

Now recall that the RV $\Delta_i \in [0,1]$ is by definition equal to the fraction of time in the interval $[0, L]$ where $Q(t) = 1$, for user $i$.[17] To compute $\mathbb{E}\{\Delta^{1/b}\}$, we define four events corresponding to the possible state transitions in the interval $[0, L]$: $\mathcal{E}_{k,l} \triangleq \{Q(0) = k \wedge Q(L) = l\}$, where $k \in \{0,1\}, l \in \{0,1\}$.

[17]Since the RVs $\Delta_i$ are IID for different users $i$, we drop for convenience the subscript $i$ in the remainder of the proof.

The probability of the events $\mathcal{E}_{k,l}$ can be written as

$$\mathbb{P}\{\mathcal{E}_{0,0}\} = \frac{e^{-\lambda_\text{p}L}}{1 + \lambda_\text{p}L}, \tag{37}$$

$$\mathbb{P}\{\mathcal{E}_{0,1}\} = \mathbb{P}\{\mathcal{E}_{1,0}\} = \frac{1 - e^{-\lambda_\text{p}L}}{1 + \lambda_\text{p}L}, \tag{38}$$

$$\mathbb{P}\{\mathcal{E}_{1,1}\} = \frac{\lambda_\text{p}L + e^{-\lambda_\text{p}L} - 1}{1 + \lambda_\text{p}L}. \tag{39}$$

Let $T_1$ denote an exponential RV with mean $\frac{1}{\lambda_\text{p}L}$, and $U_1$ denote a uniform RV in the interval $[0,1]$, independent of $T_1$. Then, we can express $\mathbb{E}\{\Delta^{1/b}\}$ as

$$\mathbb{E}\{\Delta^{1/b}\} = \sum_{k,l} \mathbb{E}\{\Delta^{1/b} | \mathcal{E}_{k,l}\} \cdot \mathbb{P}\{\mathcal{E}_{k,l}\}$$
$$= 2\mathbb{E}\{(1-T_1)^{1/b} | \mathcal{E}_{0,1}\} \cdot \mathbb{P}\{\mathcal{E}_{0,1}\} \tag{40}$$
$$+ \mathbb{E}\{(1-T_1)^{1/b} | \mathcal{E}_{1,1}\} \cdot \mathbb{P}\{\mathcal{E}_{1,1}\}.$$

The corresponding conditional PDFs can be determined as

$$f_{T_1|\mathcal{E}_{0,1}}(x) = \frac{\lambda_\text{p}L}{1 - e^{-\lambda_\text{p}L}} e^{-\lambda_\text{p}Lx}, \; 0 \le x \le 1,$$

$$f_{T_1|\mathcal{E}_{1,1}}(x) = \frac{(\lambda_\text{p}L)^2}{\lambda_\text{p}L + e^{-\lambda_\text{p}L} - 1}(1-x)e^{-\lambda_\text{p}Lx}, \; 0 \le x \le 1.$$

Using these PDFs and (37)-(39), we can expand (40) to obtain the desired result in (26). This concludes the proof.

## APPENDIX C
## DERIVATION OF $p_\text{A}$ IN (14)

Let $\nu = \frac{P^* r_0^{2b} m}{P_0} L$ with $L = e^{-2\sigma G_0}$, so that (13) can be written as

$$p_\text{A} = 1 - \frac{1}{\Gamma(m)} \mathbb{E}_{G_0} \{\gamma_\text{inc}(m, \nu)\}. \tag{41}$$

For integer $m$, we can express $\gamma_\text{inc}(m,\nu)$ in closed form [34], and (41) becomes

$$p_\text{A} = 1 - \frac{(m-1)!}{\Gamma(m)}\left(1 - \sum_{k=0}^{m-1} \mathbb{E}_\nu\left\{\frac{\nu^k e^{-\nu}}{k!}\right\}\right),$$

$$= 1 - \frac{(m-1)!}{\Gamma(m)}\left(1 - \sum_{k=0}^{m-1} \frac{(-1)^k}{k!} \left.\frac{d^k \phi_\nu(s)}{ds^k}\right|_{s=1}\right). \tag{42}$$

Approximating the MGF of the log-normal RV $L$ by a Gauss-Hermite series [35], we can write

$$\phi_\nu(s) = \phi_L\left(\frac{P^* r_0^{2b} m}{P_0} s\right)$$
$$\approx \frac{1}{\sqrt{\pi}} \sum_{n=1}^{N_\text{p}} H_{x_n} \exp\left(-\frac{P^* r_0^{2b} m}{P_0} e^{2\sqrt{2}\sigma x_n} s\right),$$

where $x_n$ and $H_{x_n}$ are, respectively, the zeros and the weights of the $N_\text{p}$-order Hermite polynomial. Consequently,

$$\frac{d^k \phi_\nu(s)}{ds^k} = \frac{1}{\sqrt{\pi}} \sum_{n=1}^{N_\text{p}} H_{x_n} \left(-\frac{P^* r_0^{2b} m}{P_0} e^{2\sqrt{2}\sigma x_n}\right)^k$$
$$\times \exp\left(-\frac{P^* r_0^{2b} m}{P_0} e^{2\sqrt{2}\sigma x_n} s\right).$$

Defining $\nu_2 = \frac{P^* r_0^{2b} m}{P_0} e^{2\sqrt{2}\sigma x_n}$, we obtain

$$\left.\frac{d^k \phi_\nu(s)}{ds^k}\right|_{s=1} = \frac{1}{\sqrt{\pi}} \sum_{n=1}^{N_P} H_{x_n}(-\nu_2)^k e^{-\nu_2},$$

which replaced into (42) leads to

$$p_A \approx 1 - \frac{(m-1)!}{\Gamma(m)} \left(1 - \frac{1}{\sqrt{\pi}} \sum_{k=0}^{m-1} \sum_{n=1}^{N_P} H_{x_n} \frac{\nu_2^k e^{-\nu_2}}{k!}\right).$$

This is the result in (14) and the derivation is complete.

## APPENDIX D
## DERIVATION OF $\mathbb{P}\{\mathsf{SINR} \geq \theta^*\}$ IN (30)

For integer $m$, and defining $\nu_3 = \frac{r_0^{2b} \theta^* m}{P_0}$, we can write (29) as

$\mathbb{P}\{\mathsf{SINR} \geq \theta^*\} =$

$$= 1 - \frac{(m-1)!}{\Gamma(m)} \left(1 - \sum_{k=0}^{m-1} \mathbb{E}_I \left\{\frac{[\nu_3(I+N)]^k e^{-\nu_3(I+N)}}{k!}\right\}\right),$$

$$= 1 - \frac{(m-1)!}{\Gamma(m)}$$
$$\times \left(1 - \sum_{k=0}^{m-1} \frac{\nu_3^k}{k!} \mathbb{E}_I\left\{\sum_{j=0}^{k} \binom{k}{j} I^j N^{k-j} e^{-\nu_3(I+N)}\right\}\right),$$

$$= 1 - \frac{(m-1)!}{\Gamma(m)}$$
$$\times \left(1 - \sum_{k=0}^{m-1}\sum_{j=0}^{k} \binom{k}{j} \frac{\nu_3^k}{k!} N^{k-j} e^{-\nu_3 N} \mathbb{E}_I\left\{I^j e^{-\nu_3 I}\right\}\right),$$

$$= 1 - \frac{(m-1)!}{\Gamma(m)}$$
$$\times \left(1 - \sum_{k=0}^{m-1}\sum_{j=0}^{k} \frac{(-\nu_3)^j}{j!} \frac{(\nu_3 N)^{k-j} e^{-\nu_3 N}}{(k-j)!} \left.\frac{d^j \phi_I(s)}{ds^j}\right|_{s=\nu_3}\right),$$

which is the result in (30). We now determine the MGF $\phi_I(s)$ of the aggregate interference power $I$. For the case of path loss and *Nakagami-m* fading, we have that $K = 1$ and $Z_{0,1} = \alpha_0^2$, where $\alpha_0^2 \sim \mathcal{G}(m, \frac{1}{m})$. In this scenario, the distribution of $I$ in (18) reduces to

$$I \sim \mathcal{S}\left(\alpha = \frac{1}{b}, \beta = 1,\right.$$
$$\left.\gamma = \pi \lambda C_{1/b}^{-1} P_I^{1/b} \mathbb{E}\{\Delta_i^{1/b}\} \frac{\Gamma\left(m + \frac{1}{b}\right)}{m^{1/b} \Gamma(m)}\right). \tag{43}$$

The corresponding MGF $\phi_I(s)$ can be determined using the following theorem.

*Proposition D.1:* The MGF $\mathbb{E}\{e^{-sX}\}$, $s \geq 0$, of the RV $X \sim \mathcal{S}(\alpha, \beta = 1, \gamma)$, $0 < \alpha \leq 2$, is given by

$$\mathbb{E}\{e^{-sX}\} = \begin{cases} \exp\left(-\frac{\gamma}{\cos\left(\frac{\pi\alpha}{2}\right)} s^\alpha\right), & \alpha \neq 1, \\ \exp\left(\frac{2}{\pi} \gamma s \ln s\right), & \alpha = 1. \end{cases}$$

*Proof:* See [37, Proposition 1.2.12]. □

From the proposition, it follows directly that

$$\phi_I(s) = \exp\left(-\frac{\pi \lambda C_{1/b}^{-1} P_I^{1/b} \Gamma\left(m + \frac{1}{b}\right) \mathbb{E}\{\Delta_i^{1/b}\}}{m^{1/b} \Gamma(m) \cos\left(\frac{\pi}{2b}\right)} s^{1/b}\right)$$

for $s \geq 0$, which is the result in (31). This concludes the derivation.

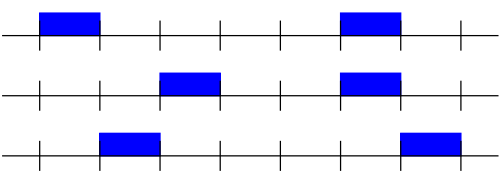

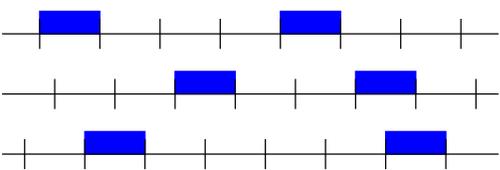

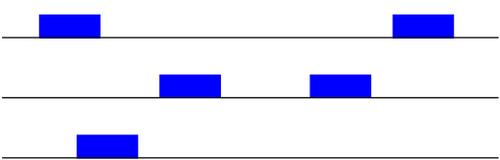